\documentclass[twocolumn,
preprintnumbers,amsmath,amssymb,prl]{revtex4}
\usepackage{amsmath,amssymb}%,amsthm}
\usepackage{dcolumn}% Align table columns on decimal point

\newcommand {\beq}{\begin{eqnarray}}
\newcommand {\eeq}{\end{eqnarray}}
\def\vev#1{\langle#1\rangle}
\def\tr{\mathop{\mathrm{tr}}}
\newcommand{\cN}{\ensuremath{\mathcal{N}}}
\newcommand{\bZ}{\ensuremath{\mathbb{Z}}}
\newcommand{\IR}{\ensuremath{\text{IR}}}
\newcommand{\UV}{\ensuremath{\text{UV}}}

\begin{document}
%%%%%%%%%%%%%%%%%%%%%%%%%%%%%%%%%%%%%%%%%%%%%%%%%%%%%%%%%%%%%%%%%%%%%%%%
%%%%%%%%%%%%%%%%%%%%%%%%%%%%%%%%%%%%%%%%%%%%%%%%%%%%%%%%%%%%%%%%%%%%%%%%

\title{27/32}% Force line breaks with \\

\author{Yuji Tachikawa}
\author{Brian Wecht}
\affiliation{School of Natural Sciences, Institute for Advanced Study, Princeton,
NJ 08450, USA}

\date{\today}% It is always \today, today,

%-----------------------------------------
\begin{abstract}
%We demonstrate that a seeming numerical coincidence is in fact representative of underlying physics.
%In particular, 
We show that when an $\cN =2$ SCFT flows to an $\cN=1$ SCFT via giving a mass to the adjoint chiral superfield in a vector multiplet with marginal coupling, the central charges $a$ and $c$ of the $\cN=2$ theory are related to those of the $\cN=1$ theory  by a universal linear transformation. 
In the large $N$ limit, this relationship implies that the central charges obey $a_\IR /a_\UV =c_\IR /c_\UV= 27/32$. This gives a physical explanation to many examples of this number found in the literature, and also suggests the existence of a flow between some theories not previously thought to be connected.
\end{abstract}

\maketitle

%%%%%%%%%%%%%%%%%%%%%%%%%%%%%%%%%%%%%%%%%%%%%%%%%%%%
%\section{Introduction}

\noindent
{\bf Introduction}:~
Although once believed to be exotic, interacting superconformal fixed points are now thought to be ubiquitous infrared  phases of four-dimensional supersymmetric gauge theories. The study of superconformal field theories (SCFTs) has seen a tremendous amount of progress over the past ten  years or so, mainly due to the impetus provided by the AdS/CFT correspondence \cite{Maldacena:1997re}, which prominently features a nontrivial superconformal theory on one side of the duality. However, such SCFTs are also interesting purely from a field theory perspective, since the large amount of symmetry present gives us access to many exact results. 

One defining property of a conformal field theory is the {\it central charge}. Two-dimensional conformal field theories have one central charge $c$, which can be found (for example) via the trace anomaly $\vev{T^\mu_\mu} = -\frac{c}{12}R$, where $T_{\mu \nu}$ is the stress-energy tensor and $R$ is the Ricci scalar curvature of the background spacetime. The central charge $c$ is known to obey a monotonicity theorem \cite{Zamolodchikov:1986gt}, which guarantees that it always decreases during renormalization group (RG) flow from the ultraviolet (UV) to the infrared (IR), $c_\IR < c_\UV$. Four-dimensional SCFTs have two central charges, denoted $a$ and $c$, which show up via the four-dimensional trace anomaly
\begin{equation}
\vev{T_\mu^\mu} = {c\over 16\pi^2}(\text{Weyl})^2 - {a\over 16 \pi^2}(\text{Euler}),
\end{equation}
where
\begin{eqnarray}
({\rm Weyl})^2 &=& R^2_{\mu\nu\rho\sigma}-2R^2_{\mu\nu}+{1\over 3} R^2,\\
({\rm Euler})&=& R^2_{\mu\nu\rho\sigma}-4R^2_{\mu\nu}+ R^2.
\end{eqnarray}
Although it was long believed that the central charge $a$ would monotonically decrease during RG flow \cite{Cardy:1988cwa},  a counterexample was recently found for $\cN=2$ SCFTs \cite{Shapere:2008un}. 

In this note, we prove a general result about RG flows between four-dimensional $\cN =2$ and $\cN = 1$ superconformal theories. In particular, we show that when an $\cN=2$ theory is deformed by a mass term for the adjoint $\cN=1$ chiral superfield inside an $\cN=2$ vector multiplet whose coupling constant is exactly marginal, 
the central charges of the infrared $\cN=1$ theory are related to the central charges of the ultraviolet $\cN=2$ theory by a simple linear transformation. If the theories have holographic duals, $a=c$ in the large $N$ limit, and our relationship implies $a_\IR /a_\UV =c_\IR /c_\UV= 27/32$. As a test of this simple general relationship, we recall many examples of such RG flows that agree with this result.
We will see that this particular number $27/32$ was encountered repeatedly
in the literature, but was never understood as a universal consequence of the flow.

%%%%%%%%%%%%%%%%%%%%%%%%%%%%%%%%%%%%%%%%%%%%%%%%%%%%%%%%%%%%
%\section{R-symmetries in $\cN=2$ and $\cN=1$ SCFTs}

\medskip
\noindent
{\bf R-symmetries in $\cN=2$ and $\cN=1$ SCFTs}:~
Our central tool for computing central charges will be the $U(1)_R$ symmetry. In particular, we will make use of the relationships \cite{Anselmi:1997am}
\begin{equation}
a=\frac{3}{32}\left[3\tr R^3-\tr R\right],\ 
c=\frac{1}{32}\left[9\tr R^3 - 5\tr R\right], \label{zot}
\end{equation} where
$R$ in this expression is the matrix of R-charges of Weyl fermions in the theory; in a strongly coupled theory $\tr R^3$ and $\tr R$ should be understood as the corresponding anomaly coefficients.
They allow us to compute the central charges via global anomalies of the $\cN=1$ R-symmetry. 
For a generic gauge theory with gauge group $G$ and matter fields $\phi_i$ in representations ${\bf r_i}$, these global anomalies are
\begin{eqnarray}
\tr R &=& |G| + \sum_i |{\bf r_i} | (R(\phi_i)-1), \nonumber \\
 \tr R^3 &=& |G| + \sum_i |{\bf r_i} | (R(\phi_i)-1)^3.
 \label{rcharges}
\end{eqnarray}
The first term in these expressions is from the gauginos $\lambda$, which have unit R-charge. The second term is from the matter fermions, which have R-charge one less than their scalar superpartners. Although (\ref{rcharges}) is written as if we have a weakly coupled description of the theory, we can often  compute these global anomalies for a strongly coupled fixed point via 't Hooft anomaly matching.

Generically, finding the superconformal R-charges in (\ref{rcharges}) is a difficult task. In $\cN=1$ theories we can use $a$-maximization \cite{Intriligator:2003jj}, although this technique only works if there are no accidental symmetries in the IR. In $\cN=2$ theories, we can use the methods of \cite{Shapere:2008zf}. Note that the R-symmetry of an $\cN=2$ SCFT is $SU(2)\times U(1)$, and the R-charges in (\ref{zot}) are those of the $\cN=1$ subalgebra. These are given by the linear combination
\beq
R_{\cN=1}=\frac13 R_{\cN=2} + \frac43 I_3
\label{rntwo}
\eeq
where $R_{\cN=2}$ is the $\cN=2$ $U(1)$ R-symmetry
and $I_a$, ($a=1,2,3$) are the $\cN=2$ $SU(2)$ R-symmetry generators.

As a quick example, consider the $\cN=2$ vector multiplet. The fields in this multiplet are the gauge boson $A_\mu$, the gaugino $\lambda_\alpha$, an adjoint fermion $\psi_\alpha$, and an adjoint complex scalar $\phi$. The fermion and the gaugino transform in a doublet of the $SU(2)$ R-symmetry, with the gaugino as the top component. The vector and the scalar are singlets under this $SU(2)$. The $\cN=2$ $U(1)$ charges are $0$ for the vector, $1$ for the gaugino and adjoint fermion, and $2$ for the scalar. Thus, we see that the $\cN=1$ R-charges work out as expected: 0 for the vector, 1 for the gaugino, $-1/3$ for the fermion, and 2/3 for the scalar. One can easily check that the charges for the $\cN=2$ hypermultiplet work similarly.

%%%%%%%%%%%%%%%%%%%%%%%%%%%%%%%%%%%%%%%%%%%%%%%%%%%%%%%%%%%%
%\section{UV and IR R-symmetries}

\medskip
\noindent
{\bf UV and IR R-symmetries}:~
Consider now an $\cN=2$ gauge theory for which all gauge coupling constants are marginal.
It can be an ordinary theory with vector and hypermultiplets,
or one in which vector multiplets couple to another strongly-coupled isolated SCFT,
as in \cite{Argyres:2007cn}.
We  deform this theory via a mass term $\delta W = m \tr \Phi^2$ for the $\cN=1$ adjoint chiral superfield $\Phi$ inside the $\cN=2$ vector multiplet. 
In the IR the theory is expected to flow to an $\cN=1$ SCFT,
with quartic superpotential generated through the decoupling of $\Phi$.

The simplest example is the flow from an $\cN=2$ theory with gauge group $SU(N_c)$
and $N_f=2N_c$ quarks $Q$, $\tilde Q$ in the fundamental representation. The IR endpoint is an $\cN=1$ theory with $SU(N_c)$ gauge group and $N_f=2N_c$ quarks, with the quartic superpotential $W\propto \tr Q\tilde Q Q\tilde Q$,
which is self-dual under Seiberg duality \cite{Seiberg:1994pq}.
Having this example in mind will help understand the argument below,
but our derivation  does not depend on a specific choice for
the Lagrangian of the UV theory. In fact, we do not even require 
a Lagrangian description of the UV theory; 
all we need is to know the 't Hooft anomalies in the UV.

We assume that there is no accidental chiral $U(1)$ symmetry in the infrared.
Then the $U(1)$ $R$-symmetry $R_\IR$ 
in the infrared $\cN=1$ theory is a linear combination of
the symmetries in the UV $\cN=2$ theory. 
The flavor symmetries of the UV $\cN=2$ theory
do not contribute to infrared R-symmetry because they are non-chiral \cite{Intriligator:2003jj}.
Thus $R_\IR$ should be the linear combination of $R_\UV$ and $I_3$ which remains
unbroken by the mass term of the adjoint $\cN=1$ chiral superfield:
\begin{equation}
R_\IR= \frac12 R_{\cN=2}+ I_3 . \label{foo}
\end{equation}  
The anomaly of $R_\IR$ is reliably calculable in the UV using the 
anomaly matching argument by 't Hooft.

We can now insert (\ref{foo}) into (\ref{zot}) to obtain the central charges in the IR in terms of those in the UV. We first calculate the global anomalies
\begin{equation}
\tr R^3_\IR= 12a_\UV-9c_\UV ,\quad 
\tr R_\IR =  24(a_\UV-c_\UV). \label{baz}
\end{equation} 
In calculating (\ref{baz}) we have used 
\begin{eqnarray}
\tr R_{\cN=2}^3 &=& \tr R_{\cN=2} = 48(a_\UV-c_\UV), \nonumber \\
\tr R_{\cN=2} I_a I_b &=& \delta_{ab} (4a_\UV-2c_\UV),
\end{eqnarray}
which are true in any $\cN=2$ SCFT (see \cite{Shapere:2008zf,Kuzenko:1999pi} for details). Additionally, any traces with odd powers of $I_3$ automatically vanish by symmetry.

We can now use (\ref{baz}) and (\ref{zot}) to give the central charges in the IR,
 \begin{align}
a_\IR&= \frac{9}{32}\left[ 4a_\UV-c_\UV \right], \\
c_\IR&= \frac{1}{32}\left[ -12a_\UV+39c_\UV \right].
\end{align}
It is easily seen that the flow from the $\cN=2$ $SU(N_c)$ theory with $N_f=2N_c$ flavors
satisfies this relation.

A general consequence of holography is that $a=c$ to leading order in $N$. Thus, for theories with holographic duals, we find, in the large $N$ limit,
\beq
\frac{a_\IR}{a_\UV}= \frac{c_\IR}{c_\UV} = \frac{27}{32} \label{main}.
\eeq
This is our main result.

%%%%%%%%%%%%%%%%%%%%%%%%%%%%%%%%%%%%%%%%%%%%%%%%%%%%%%%%%%%%
%\section{Examples}

\medskip
\noindent
{\bf Examples}:~
We now list some flows which obey (\ref{main}). We will stick to summarizing the basic features of these flows, and quoting the central charges from the literature.

\smallskip
\noindent {\bf $S^5/\bZ_2$ to $T^{1,1}$:} Consider a stack of $N$ D3-branes at the tip of a Calabi-Yau cone $X_6$. The base of this cone is a five-dimensional Einstein-Sasaki manifold $H_5$. When $H_5=S^5$, the AdS/CFT correspondence tells us that type IIB string theory on $AdS_5 \times S^5$ is dual to $SU(N)$ $\cN=4$ super Yang-Mills.

 We can break some supersymmetry by orbifolding the Calabi-Yau by any discrete $\Gamma \subset SU(3)$; this induces an orbifold action on the base. Generically such an orbifold preserves $\cN=1$ SUSY, but in the special case $\Gamma = \bZ_2$, we preserve $\cN=2$. 
The field theory is an $SU(N)\times SU(N)$ $\cN=2$ gauge theory with two hypermultiplets transforming as bifundamentals. 
We can break SUSY to $\cN=1$ by introducing a mass for each adjoint. The resulting low-energy theory and its holographic dual were found in \cite{Klebanov:1998hh}. On the gravity side, the base of the cone is now $H_5=T^{1,1}$, and the resulting Calabi-Yau is the conifold. 

On general grounds, we expect that $a=c \sim 1/{\rm Vol}(H_5)$ for any given holographic theory. As a result, we should find that $a_\IR/a_\UV = {\rm Vol}(S^5/\bZ_2)/{\rm Vol}(T^{1,1})$. The volume of $T^{1,1}$ was computed in \cite{Gubser:1998vd}, where it was found that $ {\rm Vol}(S^5/\bZ_2)/{\rm Vol}(T^{1,1}) = 27/32.$

In \cite{Gubser:1998ia,Bergman:2001qi}, this analysis was generalized by considering flows from 
arbitrary $\cN=2$ orbifolds $S^5/\Gamma$, where $\Gamma\subset SU(2)$. The mass deformation then generates a flow to theories dual to $AdS_5 \times M_\Gamma$, where $M_\Gamma$ 
is the base of a so-called ``generalized conifold." The ratio 27/32 was found both in the field theory side and in the gravity side. The same conclusion was reached in \cite{Corrado:2004bz}.

\smallskip
\noindent {\bf Flows from $\cN=4$:} In \cite{Freedman:1999gp}, the authors constructed a flow from $\cN=4$ SYM to $\cN=1$ by introducing a mass for one of the $\cN=1$ adjoint superfields. In particular, they constructed five-dimensional kink solutions which interpolate between these two theories on the gravity side. Since $\cN=4$ theories also have $\cN=2$ SUSY, our argument should apply. 

Indeed, the authors of \cite{Karch:1999pv,Freedman:1999gp} found that $a_\IR/a_\UV = 27/32$ from field theory. An earlier computation of the central charges at the endpoints of the flow was done in \cite{Khavaev:1998fb}, with the same result for the ratio of the central charges.
The $\bZ_n$ orbifold of this flow was analyzed via five-dimensional supergravity in \cite{Corrado:2002wx} where the ratio $27/32$ was found again. 
There, it was argued that there is a continuous family of flows
connecting this solution to the generalized conifold theory described above.
Such a family for $\bZ_2$ was explicitly constructed as a solution of ten-dimensional supergravity in \cite{Halmagyi:2005pn}.

\smallskip
\noindent {\bf Maldacena-Nu\~nez solutions:} In \cite{Maldacena:2000mw}, Maldacena and Nu\~nez constructed holographic theories by wrapping M5-branes on genus $g$ Riemann surfaces $\Sigma_g$. In the low-energy limit, these result in four-dimensional SCFTs with either $\cN=2$ or $\cN=1$ SUSY, depending on the embedding of $\Sigma_g$. They found that the central charges in these theories satisfy
\beq
a = \frac83 N^3 (g-1) A,
\eeq
where $A$ is a number that depends on the embedding of  $\Sigma_g$. For cases that preserve $\cN=2$ in four dimensions, they found $A_{\cN=2} = 1/8$. For $\cN=1$ embeddings, $A_{\cN=1} = 27/256$. 
Given our result, this strongly suggests that the field theory dual to the $\cN=1$ solution is the mass-deformed version of the field theory dual to the $\cN=2$ one, constructed in \cite{Gaiotto:2009gz}.
More details will be presented elsewhere \cite{InProgress}.

\medskip

We cannot resist pointing out that 27/32 occurs in an apparently unrelated context:
the value of $J^2/M^3$ for the minimally-rotating black ring was found to be
27/32 times that of the maximally-rotating black hole in five dimensions \cite{Elvang:2003yy,Emparan:2004wy}.  This same ratio also appeared in the context of graph theory in \cite{JacksonSokal}\footnote{We thank Prof. Malek Abdesselam for pointing us to this paper.}.

%%%%%%%%%%%%%%%%%%%%%%%%%%%%%%%%%%%%%%%%%%%%%%%%%%%%%%%%%%%%
%\section{Conclusions}

\medskip
\noindent
{\bf Conclusions}:~
In this brief letter, we have explained why what previously appeared to be a numerical coincidence is in fact a generic consequence of flowing from $\cN=2$ to $\cN=1$ via an adjoint mass deformation. In some cases, such as the Maldacena-Nu\~nez solutions, the ratio of central charges we discuss here had not previously been noticed. As such, we provide yet another shining jewel for the crown of numerology.

%%%%%%%%%%%%%%%%%%%%%%%%%%%%%%%%%%%%%%%%%
%\section*{Acknowledgments}

\medskip
\noindent{\bf Acknowledgments}:~
The authors thank F. Benini, D. Gaiotto,  N. Halmagyi, C. Herzog, Y. Ookouchi, and J. M. Maldacena for helpful discussions.
YT is supported in part by the NSF grant PHY-0503584, and by the Marvin L. 
Goldberger membership at the Institute for Advanced Study. 
BW is supported in part by DOE grant DE-FG02-90ER40542, and by the Frank and Peggy Taplin Membership at the Institute for Advanced Study.

%%%%%%%%%%%%%%%%%%%%%%%%%%%%%%%%%%%%%%%%%%%%%%%%%%%%%%%%%%%%%%%%%%%%%%%%%%%%%%%%%%

\end{document}